\documentclass[prl,twocolumn,showpacs]{revtex4}

\begin{document}

\newcommand*{\be}{\begin{equation}}
 
\newcommand*{\ee}{\end{equation}}

\title{Exploding Bose-Einstein condensates and collapsing neutron stars 
driven by critical magnetic fields}

\author{H. P\'erez Rojas$^{a}$, A. P\'erez Mart\'{\i}nez$^{a}$ and H. J. 
Mosquera Cuesta$^{b}$}

\address{$^{a}$ Grupo de F\'{\i}sica Te\'orica, ICIMAF, Calle E No. 309, 
10400 La Habana, Cuba \\
$^{b}$ Centro Brasileiro de Pesquisas F\'{\i}sicas, Laborat\'orio de 
Cosmologia e F\'{\i}sica Experimental de Altas Energias \\ 
Rua Dr. Xavier Sigaud 150, CEP 22290-180, Urca, Rio de Janeiro, RJ, Brazil}

\date{\today}

\begin{abstract}
The problem of a condensate of a relativistic neutral vector boson gas
constituted of particles bearing a magnetic moment is discussed. Such a
vector boson system is expected to be formed either by parallel
spin-pairing of neutrons in a sufficiently strong magnetic field, or by
neutral atoms under specific conditions of magnetic field strength and
density. A strong self-magnetization arises due to a Bose-Einstein-like
condensation. Then the system, which may resemble the superfluid said
to exist in the core of neutron stars, becomes more unstable under
transverse collapse than the ordinary fermion gas. In the
nonrelativistic limit of laboratory conditions, an analogy with the
behavior of exploding Bose-Einstein condensates for critical values of
magnetic field strength and particle density; reported by several authors, 
is briefly discussed.
\end{abstract}


\maketitle

\section{Introduction}

The investigation of the properties of a condensate of vector neutral
bosons having a magnetic moment is of very  wide interest, its
applications ranging from condensed matter, i. e., Bose-Einstein
Condensates (BECs), to astrophysics, e. g., neutron stars. Here we
propose that a vector condensate arises from the Bose particles
resulting from the coupling of spin up paired neutral particles. This
would lead to an approximate model for ferromagnetic coupling among
neutrons, which is expected to take place in presence of extremely strong
magnetic fields. Because of this special configuration one expects that
the resulting Bose particle to have a mass of order twice the mass of
the fermion originating it.

The idea of vector pairing, i.e. parallel spin-pairing instead of
scalar or antiparallel pairing, is justified if we start from the
essential property that the neutron ground state corresponds to spin
alignment along the external field, and for fields large enough, the
difference between parallel and antiparallel pairing is a significant
fraction of the energy of the pair. It is expected then that the spin
parallel be a more energetically favorable coupling than the
antiparallel or scalar one. Thus, the superfluid, which is the physical
realization of the vector condensate, is also a ferromagnet. For
charged particles the situation is similar, and for electrons; in
particular, this has lead to the suggestion of a re-entrant
superconductivity in some condensed matter systems immerse in magnetic
fields of order $10^6$~G. In a system as this, a large population of
parallel spin electrons is expected to occur in the Landau ground state
$n=0$ (See Boebinger and Passner \cite{Boebinger} and references quoted
therein).

\section{Particle energy spectrum}

We start by considering a relativistic neutral boson gas of particles
with nonzero magnetic moment. A variety of physical arguments allow for
the bosons to be described as a neutral vector gauge field, even if it
is the result of a pairing of fermions. Its spectrum can be derived
from the poles of the inverse propagator for a massive gauge neutral
boson that couples to an external magnetic field through the coupling
constant $g=M q$, where $M$ is the mass of the particles and $q$ has
the dimensions of charge. The inverse propagator reads

\begin{equation}
D^{-1}(k_{\mu}, F_{\mu \nu}, M)=(k^2+M^2)\delta_{\mu \nu} -
k_{\mu}k_{\nu}-ig F_{\mu \nu},
\end{equation}

\noindent where $F_{\mu \nu}= B (\delta_{\mu 1}\delta_{\nu
2}-\delta_{\mu 2}\delta_{\nu 1})$ is the electromagnetic field
tensor describing the field $B$, which is pointing along the $3$-rd
axis in this our treatment. It is easy to find the spectrum from

\be
Det D^{-1}(k_{\mu}, F_{\mu \nu}, M)=0 \; ,
\ee

 as

\begin{equation}
E(p, B) =\mp \sqrt{p_3^2+p_{\perp}^2+M^2 +\eta(\sqrt{p_{\perp}^2+M^2}) qB}
\label{spec},
\end{equation}

where $\eta=\mp 1$ and the minus and plus signs in front of the
square root indicate particles aligned respectively along and
opposite to the magnetic field $B$. The ground state corresponds
to the case $\eta=-1$. The magnetic moment of the particles is in
this case defined as

\be
{\it m} = - \frac{\partial E(0, B)}{\partial B} = \frac{q}{\sqrt{M^2-MqB}}\; .
\ee

\section{Thermodynamical properties}

Starting from Eq.(\ref{spec}), one can write an expression for the
thermodynamical potential 

\be 
\Omega = - T \ln {\cal Z}\; , 
\ee

where ${\cal Z}$ is the partition function. Once $\Omega$ is known, if 
$\mu$ is the chemical potential, the density is given by the relation
$N = - \frac{\partial \Omega}{\partial \mu}$.  Its expression reads

\begin{eqnarray}
N & = & \frac{2\pi}{\hbar^3} \Sigma_{\eta} \int_{-\infty}^{\infty} p_{\perp}
dp_{\perp} dp_3 \left[\frac{1}{e^{(E(\eta) - \mu)/T} - 1} \right. \nonumber \\
& - & \left. \frac{1}{e^{(E(\eta) + \mu)/T}-1}\right], \nonumber \label{den}
\end{eqnarray}

and the thermodynamical potential turns out to be

\begin{eqnarray}
\Omega & = & - \frac{T}{4 \pi^2} \sum_{\eta=1,-1} \int_{0}^{\infty} p_{\perp} dp_{\perp} dp_3 \left[\ln (f^{-}f^{+}) \right. \nonumber \\
& + & \left. T^{-1} E(\eta)\right], 
\label{termodp}
\end{eqnarray}

\noindent where $f^{\mp}=(1 - e^{-(E(\eta) \mp \mu)/T})$, and the
second distribution term in Eq.(\ref{den}), as well as the second
logarithmic term in Eq.(\ref{termodp}), account for the antiparticle
contribution, whereas the term $T^{-1}E(\eta)$ accounts for the quantum
vacuum term. Its contribution is independent of $T$ and is expected to
be significant for fields of order $qB \sim M$, a case we will not
consider in what follows.  Also, the antiparticle terms, which become
interesting only at very high temperatures, will not be considered in
this approach because we are focusing our attention in either the case
of a specially tuned magnetic field strength in an atomic BEC, or the
case of a very strong magnetic field in a model of a canonical neutron
star, usually said to be a ``cold" degenerate neutron gas. Then, one can
show that for fixed $N$, $\mu$ is a decreasing function of $T$. Thus,
by decreasing $T$ (or by increasing the density) $\mu$ increases, and
reaches its maximum value at the ground state energy (magnetic moment
up) $\mu=\sqrt{M^2-MqB}$, at some critical temperature $T_c$. In other
words, the neutral particle temperature Green's function has an
infrared pole at

\be
p_4=i\omega=\pm i(E(0,B)-\mu) = i(\sqrt{M^2-MqB}-\mu) = 0\; .
\ee

As in the usual theory of Bose-Einstein condensation, for
temperatures $T<T_c$, the integral $N(T)<N$ gives the number of
excited particles, the condensate being 

\be
N_0 = N-N(T) = N(1-(T/T_c)^{3/2})\; . 
\ee

Obviously, due to the magnetic field the system is subject to strong
magnetostrictive forces expressed in anisotropic pressures
\cite{landau,jackson}: they are different along and perpendicular to
${\bf B}$. The anisotropic pressures for the gas of magnetized
particles were obtained for charged particles in previous papers
\cite{Shabad,Chaichian}, and for neutral particles in a recent one
\cite{Aurora}. The pressure along the field is 

\be
P_3=-\Omega\; , 
\ee

and perpendicular to it is 

\be
P_{\perp}=-\Omega-B{\cal M}\; , 
\ee

where ${\cal M}$ is the magnetization. Such an anisotropy can be
compensated by deforming the body, but in some limits (discussed in
Refs.  \cite{Chaichian,Aurora} and extended here to BECs), it leads to
a collapse of the system of particles.

One can estimate that near the condition for condensation. By defining 
$\bar M = \sqrt{M^2-MqB}$, the density of excited particles in Eq.(\ref{den})
is approximately given by

\begin{equation}
N \sim \frac{T}{4\pi} \sqrt{{\bar M}^2-\mu^2}\; ,
\end{equation}

\noindent which decreases as $\mu \to \bar M$. The isotropic
pressure decreases more strongly, as 

\be
P_0 = -\Omega \sim \frac{T}{6\pi} \left({\bar M}^2-\mu^2\right)^{3/2}\; . 
\ee

In other words, the condensate does not contribute to the
thermodynamical potential, or what is the same, to the longitudinal
pressure. The thermodynamic potential of the condensate is thus zero,
but its (rest) energy is non-zero, since $U_0=\mu N=\bar M N$. The
condensate contributes, however, negatively to the transverse pressure,
as we shall see below.

\section{Magnetic behavior}

In general, it is expected to have a background of excited bosons and
fermions with the total transverse pressure having contributions of
both fermions and excited bosons. Meanwhile, the magnetization contains
in addition the contribution from the boson condensate, which may be
very large.  For bosons, after the phase transition to a  condensate,
the magnetization must be written as

\be
{\cal M}_b=-\partial \Omega/\partial B + N_0 m\; ,
\ee

where the second term contains the  contribution to the magnetization due
to the condensate energy

\be
N_0 \frac{\partial \bar M}{\partial B} = N_0 m\; .
\ee

In the present case, the total transverse pressure is then \cite{Chaichian,Aurora}

\be
P_{t\perp}=-(\Omega_b + \Omega_f) -B({\cal M}_b + {\cal M}_f)\;, 
\ee

where $t$ refers to total, and $b$ and $f$ refer, respectively, to
bosons and to fermions.  If there is a large amount of bosonization and
most of the particles are in the condensate, i.e., when $-(\Omega_b +
\Omega_f)$ is equal or smaller than the negative pressure term
$-B({\cal M}_b + {\cal M}_f)$, either the total pressure perpendicular
to the field vanishes or becomes negative, and therefore system should 
collapse.  We consider two cases: a) the relativistic neutron star, and 
b) the nonrelativistic gas of a few millions of atoms, that is a BEC.

\subsection{Highly magnetized neutron stars}

For a neutron star, and in the case where there is neutrons'
spin-pairing parallel to $B$, which leads to an effective spin-1 boson
particle as the one described above, having an effective mass as that
of a neutron $m_n$. Thus, even assuming temperatures of $\sim 10^{8}$K,
since $m_n/T \sim 10^{5}$, the system must be considered as highly
degenerate, that is, below the critical temperature ($T_c \geq
10^{11}$~K), and the density of excited particles is negligibly small
as compared with the condensate density. Also, the Bose-Einstein
condensate would lead to a more energetically favorable state than the
fermion gas, as is seen by starting from  the general expression for
the internal energy density,

\be
U = \mu N + TS + \Omega\; . 
\ee

As for the condensate it is $\Omega=0$, and

\be
S = - \frac{\partial \Omega}{\partial T} = 0\; ,
\ee

implying that its internal energy is 

\be
U_b\sim \mu_b N_b\; ,
\ee

where $\mu_b \leq 2M_f$ and $N_b \leq {N_f}/2$. Meanwhile, for the Fermi 
gas it is 

\be
U_f \sim \mu N_f + \Omega_f\; , 
\ee

where $\Omega_f$ is positive and of order $\mu N_f$, with $\mu_f > M_f$. 
Thus, $U_b < U_f$. As pointed out above, the Bose condensate bears the 
properties of a superfluid and also of a ferromagnet. Therefore, most 
of the system must be in the condensate state.

If initially a strong magnetic field is created locally in the star, say
on order of $B_i \sim 10^{12-13}$~G, as inherited from the implosion of the
progenitor (a supergiant red star) by the mechanism of flux conservation,
that field would be amplified by the Bose-Einstein condensation. As the
particles in the condensate are described by the spectrum of Eq.(\ref{spec}),
then in the limit of zero momentum,  and for $\eta=-1$, the condensate
becomes polarized. Thence, one can estimate the magnetization as given as
${\cal M}_f=N_0 m$ (the thermodynamical potential of the excited particles
is obviously zero).

By taking $N_0 \sim 10^{39}$, $m \sim 10^{-23}$, one gets that the
system generates a spontaneous magnetization ${\cal M}_f\sim
10^{16}$~G, leading to a self-consistent field $B_f=4\pi{\cal M}_f \sim
10^{17}$~G. In this sense, the bosonization we are describing is a
simplified version of a more realistic superfluid or ferromagnetic
coupling in which the particles interact usually with several
neighbors, whereas we are assuming in our bosonization model that it
results from two-particle vector pairing. However, by calling $B_f$,
$B_i$ the final and initial fields, respectively, we see that (in the
present case) it leads to an internal field parameter of order
$x=B_f/B_i \sim 10^{3-4}$. Then, the pressure perpendicular to ${\bf
B}$ would be negative and on order of $10^{33-34}$~erg~cm$^{-3}$, hence
driving a transversal collapse.

\subsection{The exploding BECs of atoms}

The non-relativistic limit can be discussed straightforwardly, since
the sort of condensation we are discussing is a low-momentum
phenomenon. In the case of  BECs in the laboratory various experimental
groups estimate the BECs density as around $N\sim 10^{14}$ and $m\sim
10^{-20}$.  For laboratory magnetic fields of order $10^{3}$~G, this
combination of physical properties leads to a negative pressure of
$\sim 1$d~yn~cm$^{-2}$, which means a compression of the condensate
perpendicular to the field. This compression may lead itself to an
implosion or collapse, followed by a re-expansion or explosion of the
BEC. Implosion, followed by an explosion of BECs have been observed in
laboratory experiments by several groups (see Ref.\cite{Ketterle} and
references therein).  Although such a phenomenon is attributed to an
effect produced by finely tuning the magnetic field strength so as to make the
interatomic distance to be in a region of repulsive potential, it seems
to us that the negative pressure contribution might play a significant
if not a fundamental role in driving the exploding BECs.

\section{Discussion and Conlcusion}

We have shown that the spin-up pairing of neutral particles, neutrons
for instance, to form a spin-1 vector particle in a system pervaded by
a critical magnetic field, could naturally provide an explanation for
the peculiar behavior of both strongly magnetized neutron stars and
Bose-Einstein Condensates of atoms. those systems seem to implode and 
explode under extreme conditions. Although, the physical macroscopic
properties of those systems differ by many orders of magnitudes, we
claim that the same physics seems to be operating in both of them. The
systems are driven by a concommitant action of magnetic fields and
particle densities, which under critical conditions may be responsible
for their implosion and subsequent explosion as currently observed in
laboratory systems.

\end{document}